\documentclass[twocolumn,showpacs,preprintnumbers,amsmath,amssymb]{revtex4}

\usepackage{epsfig}
\usepackage{dcolumn}
\usepackage{bm}

\begin{document}

\author{Jordi Marcos, Antoni Planes and Llu\'{\i}s Ma\~nosa}

\address{Departament d'Estructura i Constituents de la Mat\`eria, \\
Facultat de F\'\i sica, \\
Universitat de Barcelona. \\
Diagonal, 647, E-08028 Barcelona, Catalonia}

\author {Am\'{\i}lcar Labarta and Bart Jan Hattink}

\address{Departament de F\'{\i}sica Fonamental, \\
Facultat de F\'\i sica, \\
Universitat de Barcelona. \\
Diagonal, 647, E-08028 Barcelona, Catalonia}

\date{\today }

\title{Martensitic  transition  and  magnetoresistance in  a  Cu-Al-Mn
shape memory alloy. Influence of ageing.}

\begin{abstract} We have studied the effect of ageing within the
miscibility gap on the electric,    magnetic    and thermodynamic
properties    of    a non-stoichiometric   Heusler Cu-Al-Mn
shape-memory  alloy,   which undergoes a martensitic transition from a
$bcc$-based ($\beta$-phase) towards    a close-packed    structure
($M$-phase).     Negative magnetoresistance  which  shows an  almost
linear  dependence on  the square  of  magnetization  with  different
slopes  in  the  $M$-  and $\beta$-phases, was  observed.  This
magnetoresistive  effect has been associated   with  the   existence
of   Mn-rich  clusters   with  the Cu$_2$AlMn-structure.  The effect
of  an applied magnetic field on the martensitic  transition has  also
been studied.   The entropy  change between the $\beta$- and
$M$-phases shows negligible dependence on the magnetic  field but  it
decreases significantly  with annealing  time within the miscibility
gap.  Such  a decrease is due to the increasing amount   of
Cu$_2$MnAl-rich    domains   that   do   not   transform
martensitically.
\end{abstract}


\maketitle

\section{INTRODUCTION}
\label{introduction}

\noindent

The shape-memory  effect \cite{Otsuka98} is  characteristic of certain
alloys   (the  so-called   shape-memory  alloys),   which   exhibit  a
martensitic   transition   (MT)   from   an   ordered   $bcc$   -phase
($\beta$-phase)   towards   a   close-packed   low-temperature   phase
($M$-phase).   This  effect  is  related  to  unique  thermomechanical
properties  such  as  the  ability  to recover  from  large  permanent
deformations produced in the  $M$-phase by the reverse transition when
temperature is increased. During the nineties a great deal of interest
has been devoted to the study and development of magnetic shape-memory
materials.   This interest  is  mostly  due to  the  possibility of  a
magnetic  control of  the  shape-memory effect,  which  has been  made
evident  in the  ferromagnetic  Ni-Mn-Ga alloy  close  to the  Heusler
composition Ni$_2$MnGa  \cite{Ullakko96}. Furthermore, these materials
have  been  shown  to  exhibit  unexpected  pretransitional  behaviour
\cite{Zheludev95}.

The present  paper deals with the  study of the  Cu-Al-Mn alloy.  This
Hume-Rothery material \cite{Ahlers96} shares a number of features with
the  Ni-Mn-Ga alloy  system.   It displays  the same  high-temperature
crystallographic   structure,   a   martensitic  transformation   with
associated  shape-memory effect  and interesting  magnetic properties.
For  Cu-Al-Mn,  however,  the   martensitic  transition  occurs  in  a
composition  region  far from  the  Heusler  stoichiometry.  For  this
composition  range,   the  $\beta$-phase   is  only  stable   at  high
temperatures,  but can  be  retained at  low-temperature  by means  of
suitable cooling.  During this  cooling the system develops an ordered
$L2_1$  structure   ($Fm3m$,  Heusler  symmetry)   in  two  successive
disorder-order transitions: $A2$  $(Im3m)$ $\rightarrow$ $B2$ $(Pm3m)$
at $T_{c1}$ and $B2$ $\rightarrow$ $L2_1$ at $T_{c2}$ \cite{Obrado98}.
Upon  further  cooling it  undergoes  a  martensitic  transition at  a
temperature which is  strongly composition dependent.  This transition
has a diffusionless nature  which ensures that the atomic distribution
of the $L2_1$ phase is inherited by the $M$-phase.  It is worth noting
that  this feature is  common to  all Cu-based  shape-memory materials
\cite{Ahlers86}.

Magnetic  properties arise from localized magnetic moments at Mn-atoms
as  occurs  in  the  Ni-Mn-Ga system \cite{Webster88}. These magnetic
moments are coupled through an oscillating effective interaction (RKKY
interaction).   ALCHEMI   (Atom   Location   by  Channelling  Enhanced
Microanalysis) experiments \cite{Nakanishi93} have shown that Mn atoms
are  located  preferentially  in one of the four distinguishable $fcc$
sublattices  of  the  $L2_1$  structure  (the  4b  sites in Wyckoff
notation).  For this configuration, ferromagnetic coupling is dominant
and  close to the Cu$_2$AlMn composition the system is ferromagnetic.
However,  for  non-stoichiometric  alloys  the 4b sites are not fully
occupied  by  Mn-atoms  and  this results in magnetic disorder, which
gives  rise  to  different  magnetic  behaviour  depending  on  the
temperature  range  \cite{Prado98,Obrado99}.  Magnetic  clustering has
been  suggested to be at the origin of the magnetoresistive properties
recently      reported      in      Cu-Al-Mn     melt-spun     ribbons
\cite{Murthy95,Sugimoto98}.  An interesting feature is the fact that a
phase  separation between Cu$_3$Al-rich and Cu$_2$AlMn-rich phases may
occur  below  the $L2_1$ ordering line. It is therefore expected that
the  magnetic and structural properties of the system are sensitive to
the  temperature  history  of  the  material (i.e., ageing). In
particular, isothermal  annealing at a temperature within the
miscibility gap will result  in  the  growth  of  magnetic  clusters.
The existence of a miscibility  gap  was  first  reported for the
Cu$_3$Al $\rightarrow$ Cu$_2$AlMn  pseudobinary composition line
\cite{Bouchard75} and later confirmed  for  Cu-rich  systems  with  a
composition  that slightly deviates   from  this  line
\cite{Kainuma98}.  Interestingly,  this composition  range  includes
that  for  which  Cu-Al-Mn  displays a martensitic  transition.
Recently,  it  has been theoretically shown \cite{Marcos01}  that  the
influence of magnetic degrees of freedom on configurational  phase
stability  is  at  the  origin of this phase separation.

The present paper is aimed at experimentally studying the influence of
magnetism on  the martensitic  transformation in Cu-Al-Mn.   Since the
magnetic properties of  this material are expected to  be sensitive to
ageing,  this  effect  will be studied. In particular, we focus on the
study  of  magnetotransport  and  magnetic  properties  through  the
martensitic transition  of bulk  samples. The results  will contribute
to  gain a deeper understanding of the magnetoelastic interplay, which
will   be   studied  at  different  levels  of  coupling.

The paper is organized as follows.  In section \ref{experimental}, the
experimental details  are outlined.  Section  \ref{results} deals with
the   experimental   results,   which   are   discussed   in   section
\ref{discussion}.

\section{EXPERIMENTAL DETAILS}
\label{experimental}

Measurements  were performed  on  a Cu-Al-Mn  polycrystal (grain  size
$\sim 100 \mu$m) prepared by melting pure elements (99.99$\%$ purity).
The nominal  composition of the studied  alloy is Cu;  22.8 at$\%$ Al;
9.0 at  $\%$ Mn. From the ingot,  rectangular specimens (approximately
12 $\times$ 4 mm$^2$ and  0.01 mm thick for resistance measurements or
0.5 mm for calorimetric and magnetic measurements) were first cut with
a low-speed diamond saw and  mechanically polished down to the desired
thickness.   All  samples were  annealed  for 10  min  at  1080 K  and
quenched  in a mixture  of ice  and water.   This fast  cooling avoids
precipitation  of equilibrium  phases,  but enables  the two  ordering
transitions  to  the  $B_2$   and  $L2_1$  structures  to  take  place
($T_{B_2}$  = 849  K and  $T_{L2_1}$  = 795  K \cite{Obrado98}).   The
nominal martensitic  transition temperature of  as-quenched samples is
$T_M$ = 157  $\pm$ 1 K.  For the studied  composition the structure of
the $M$-phase is 18$R$  \cite{Obrado97}. This structure is monoclinic
but it is usually described   by   a   larger (approximately
orthorhombic) unit  cell containing   18   close-packed atomic planes
along the  c-axis \cite{Lovey87}.

Phase  separation between  Cu$_3$Al-rich  and Cu$_2$AlMn-rich  domains
occurs  under  very  slow  cooling  from  high  temperature.   In  the
as-quenched specimens phase separation can be induced by means of post
annealing within the miscibility gap. This process takes place at slow
rates so that it is completely negligible at room temperature (even on
a  month  time scale). In all cases, ageing consist in annealing at a
temperature   $T_a  =$  473  K  for  selected  times.  We  chose  this
temperature  because  it  is  located slightly below the limit of the
miscibility gap for the studied composition. Actually, above 525  K no
phase separation was detected, while 50  K below, the magnitude of the
observed  effects  is  similar, but they occur on a longer time scale.

Four kind of measurements  were performed: electrical resistance under
applied   magnetic   field,   magnetization,   ac-susceptibility   and
calorimetry.  Electrical resistance was measured  from 5 K up to 300 K
using  an ac four-probe method. For magnetization measurements a SQUID
magnetometer  was  used.    For  calorimetric  measurements  a  highly
sensitive and  fast response (around  20 s time constant) calorimeter,
which  was specifically designed  for the  study of  solid-solid phase
transitions, was utilized \cite{Manosa96}. Calorimetric, magnetization
and  ac-susceptibility  runs were  performed  through the  martensitic
transition in the range from  100 K to room temperature.  Calorimetric
measurements were carried  out at a rate of 0.5  K/min, and pairs of
data  (calorimetric  output and  temperature)  were  recorded every  2
seconds.  Ac-susceptibility measurements were performed at a frequency
$f$ = 66 Hz.

\section{RESULTS}
\label{results}

\begin{figure}[ht]
\begin{center}
\epsfig{file=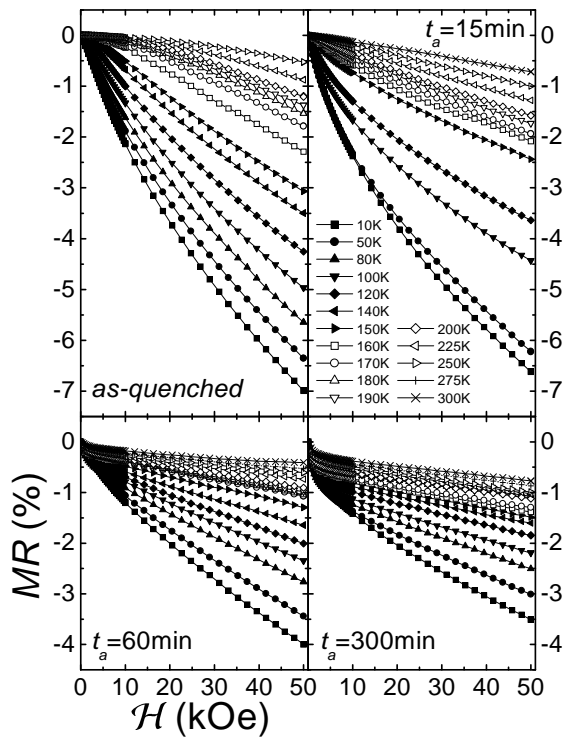, width=7.5cm}
\end{center}
\caption{Magnetoresistance  $MR$ versus magnetic  field ${\cal H}$ at
selected temperatures and for different annealing times at $T_a = 473$
K.} \label{FIG1}
\end{figure}

Electrical resistance  measurements were carried out  by first cooling
the sample down to 5 K and then increasing the temperature in steps up
to  300  K.  At  each  plateau, the  resistance  $R$  was measured  at
different values of the magnetic  field ${\cal H}$ at intervals of 345
Oe from 0 to 10 kOe, and at 2100 Oe intervals within the range from 10
to 50 kOe.  The magnetoresistance $MR$, defined as the relative change
of  $R$  with  ${\cal  H}$,  is computed  as:  $MR  =  \left[R(T,{\cal
H})-R(T,{\cal H}=0)\right]/ R(T,{\cal H}=0)$.  In Fig.\ref{FIG1}, $MR$
is plotted  as a function of  ${\cal H}$ at  selected temperatures for
samples  subjected to  increasing annealing  times at  $T_a$.   In all
cases,  the  alloy  exhibits  negative  magnetoresistance,  i.e.,  the
resistance decreases as magnetic  field increases and the magnitude of
the change is higher at  low temperature.  Actually, the maximum $MR$,
of about 7$\%$, was obtained at $T \simeq 10$ K and ${\cal H}$ = 50 kOe
for  the as-quenched  specimen. Overall,  the effect  of ageing  is to
reduce  the  $MR$ at  high  fields. However,  at  low  fields a  small
increase is  observed, which is associated with  the material becoming
magnetically softer.  For instance, the  $MR$ at $T= 300$ K and ${\cal
H}=10$ kOe increases from a practically null value for the as-quenched
state to $\sim 0.5 \%$ for the long-term annealed state.

\begin{figure}[ht]
\begin{center}
\epsfig{file=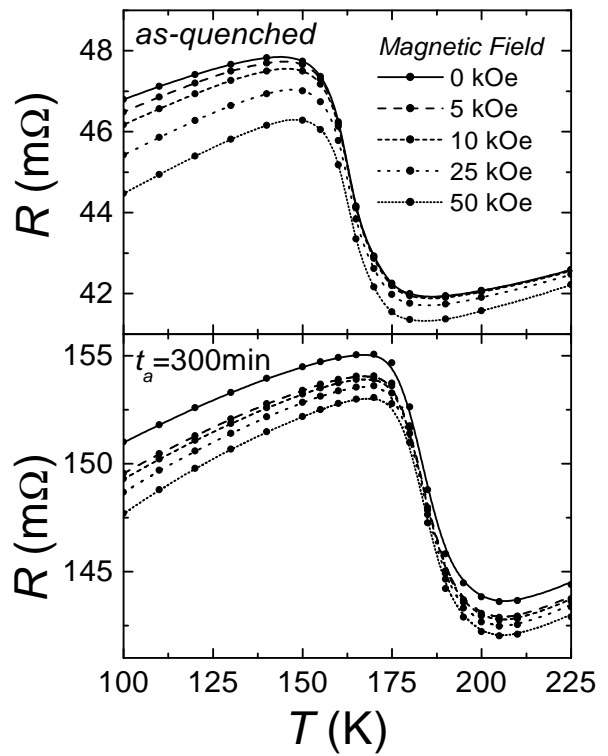, width=8cm}
\end{center}
\caption{Electrical resistance $R$  versus temperature $T$ at selected
values of the magnetic field  ${\cal H}$ for an as-quenched sample and
for a sample annealed 300min at $T_a = 473$ K .} \label{FIG2}
\end{figure}

An  interesting  feature concerns  the  influence  of the  martensitic
transition on the $MR$. This effect is better revealed by plotting the
change of  $R$ with  $T$ across  the MT at  selected values  of ${\cal
H}$. This  is  shown  in  Fig. \ref{FIG2}  for  an  as-quenched sample
and   for   a  sample  aged  for  300 min.  The  curves  obtained  are
equivalent  to those  obtained directly  by measuring  $R$  versus $T$
under continuous  heating at different  constant values of  ${\cal H}$
\cite{Marcos01b}.  The  large change in resistance  in the temperature
range between 150  K and 175 K for the  as-quenched sample and between
175 K and  200 K for the  annealed sample is due to  the occurrence of
the MT.   We will take  the temperature difference $\Delta  T$ between
the  maximum and  the  minimum of  the  $R$ versus  $T$  curves as  an
estimation of the spread  in temperature of the transition (difference
in   the  starting  and   finishing  transition   temperatures).   The
corresponding  change in  resistance will  be denoted  by  $\Delta R$.
Regardless of the  ageing influence, the effect of  the magnetic field
is mainly to decrease the  resistance of the $M$-phase (the resistance
in the  $\beta$-phase decreases by a  lower amount), which  leads to a
reduction  in $\Delta  R$ as  the  field is  increased.  In  addition,
$\Delta T$  is reduced  when the field  increases (in  the as-quenched
system this reduction amounts to $\sim$6  K for an applied field of 50
kOe).

In  order to  characterize  the  magnetic response  of  the system  we
carried out magnetization  (${\cal M}$) and ac-susceptibility ($\chi$)
measurements through  the martensitic transition.   These measurements
were undertaken on samples  subjected to increasing annealing times at
$T_a$.   Figure  \ref{FIG3}(a)  shows ${\cal  M}$ versus  $T$  curves
at selected  values of  the  magnetic field.   Figure  \ref{FIG3}(b)
shows magnetization  versus  ${\cal  H}/T$  curves.  It  is
interesting  to correlate    the   magnetization   with    the
behaviour    of   the magnetoresistance.  Figure   \ref{FIG4} shows
the $MR$ as  a function of the  square  of  magnetization  ${\cal
M}^2$ for   as-quenched and long-term  aged  specimens.   For   each
sample, all  data  for the $\beta$-phase scale  on a single  curve,
while data for  the $M$-phase scale  on another  curve.  In  the
as-quenched state, the two scaling curves  are  approximately linear,
while remarkable deviations  from linearity are  evident in the
annealed system, particularly  for high values of  ${\cal M}$.
Moreover,  in the annealed state,  the scaling (in both $\beta$- and
$M$-phases) is not as good as in the as-quenched state.

\begin{figure}[ht]
\begin{center}
\epsfig{file=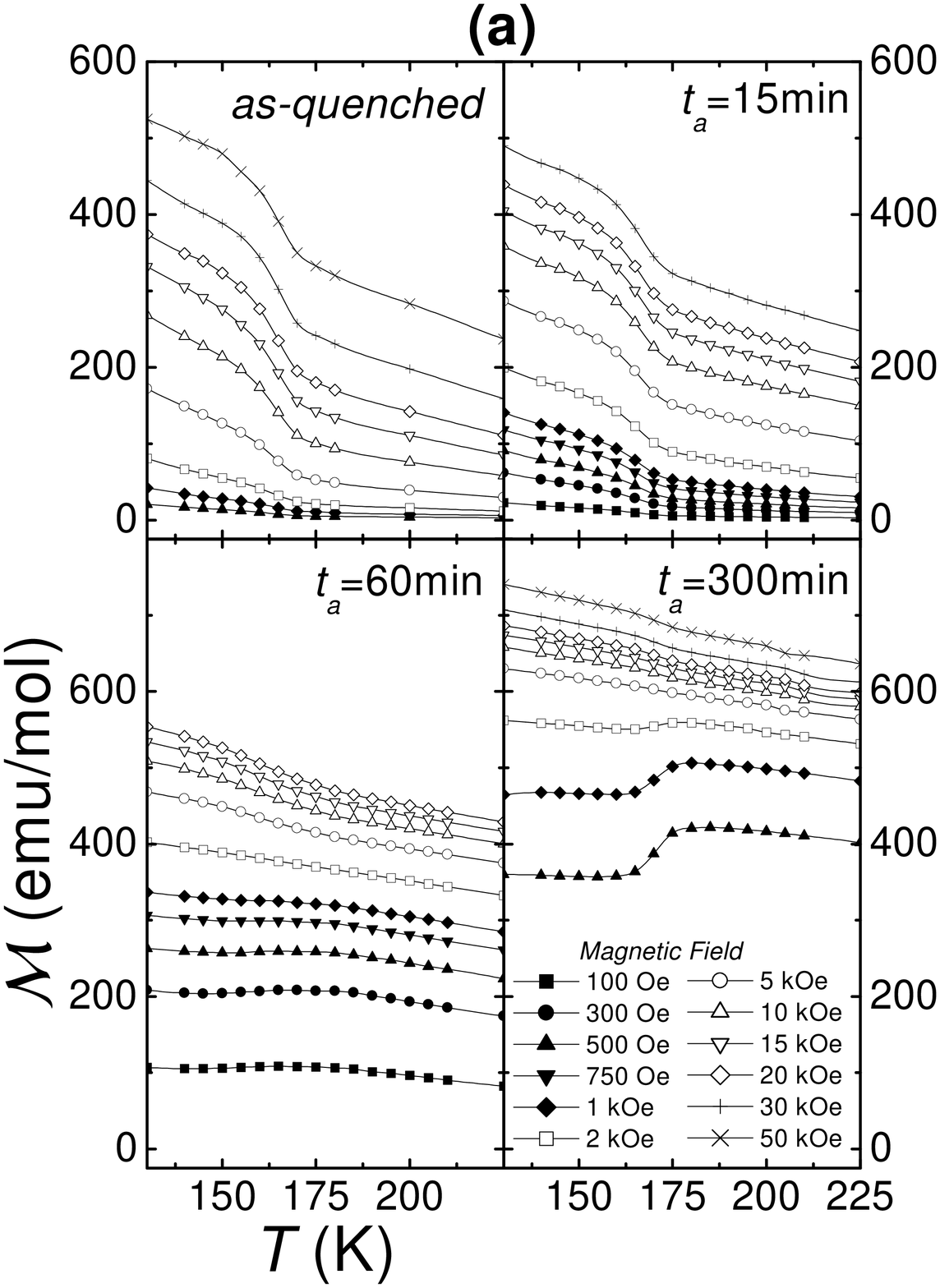, width=7.5cm}
\end{center}
\end{figure}

\begin{figure}[ht]
\begin{center}
\epsfig{file=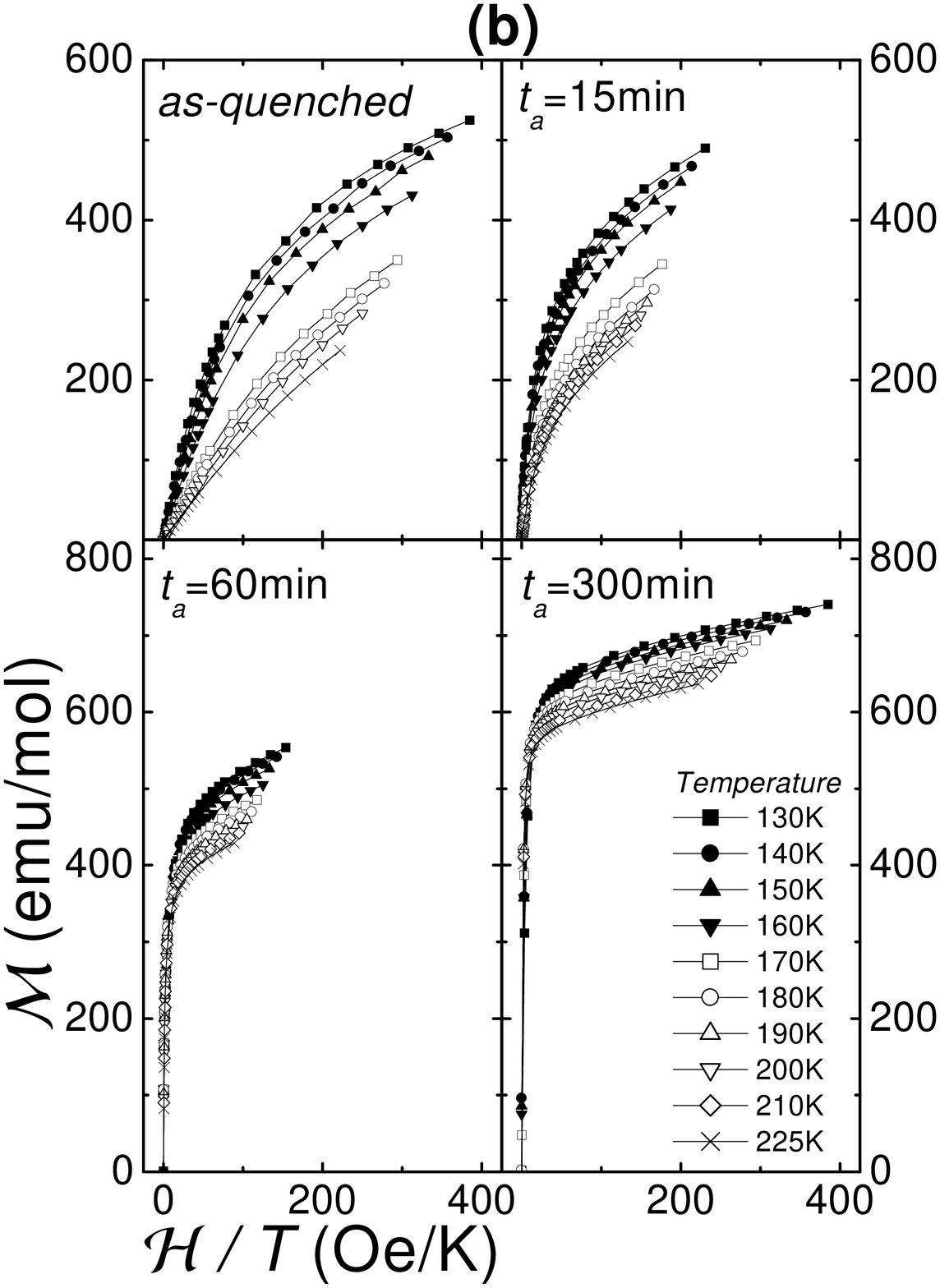,width=7.5cm}
\end{center}
\caption{(a) Magnetization  ${\cal  M}$  versus temperature $T$ at
selected values of ${\cal H}$  for different annealing times at $T_a =
473$ K. (b) Magnetization as a function of ${\cal H}/T$ for several
selected temperatures;  solid symbols correspond to  the $M$-phase and
open symbols to the $\beta$-phase.} \label{FIG3}
\end{figure}

\begin{figure} [ht]
\begin{center}
\epsfig{file=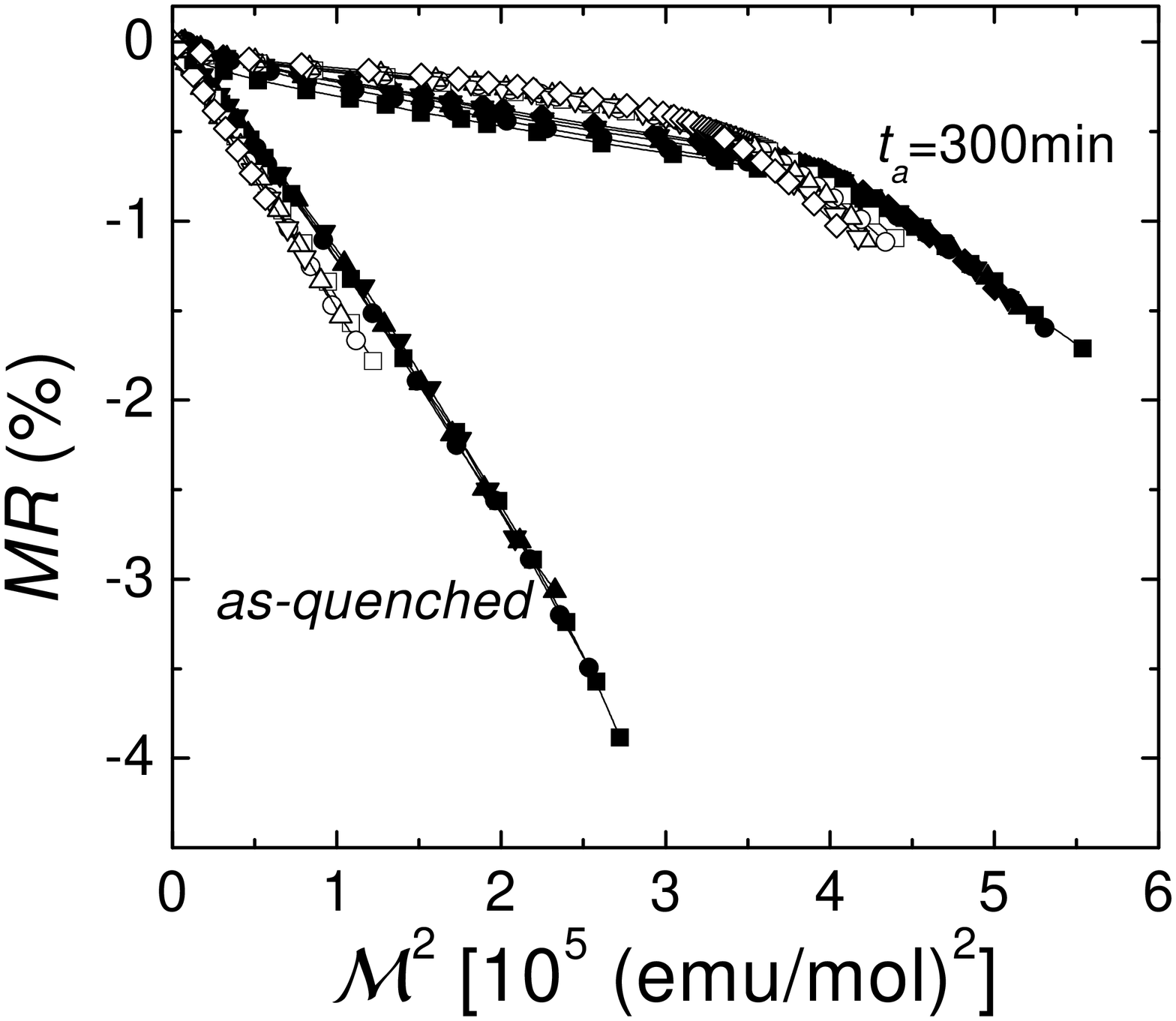, width=8cm}
\end{center}
\caption{Magnetoresistance  $MR$ as a  function of  the square  of the
magnetization.   Solid symbols  correspond to  the $M$-phase  and open
symbols,  to  the $\beta$-phase.  Results  for  an as-quenched and  a
long-term aged ($t_a$ = 300 min) specimen are shown.} \label{FIG4}
\end{figure}

\begin{figure}[t]
\begin{center}
\epsfig{file=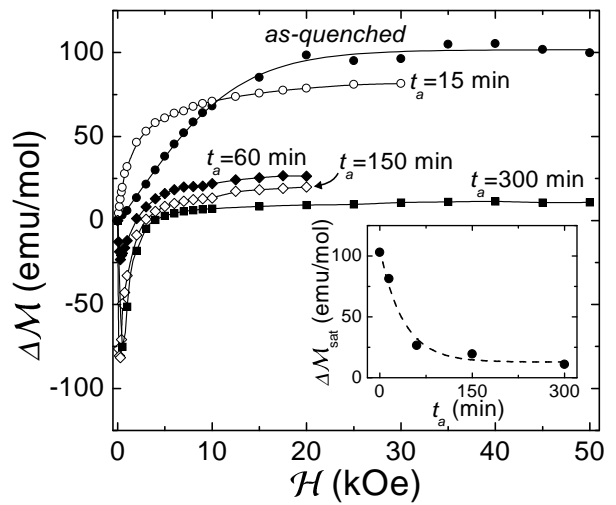, width=8cm}
\end{center}
\caption{Magnetization   change $\Delta{\cal  M}$  at the martensitic
transition   versus   magnetic   field   ${\cal  H}$ at     selected
annealing times at $T_a = 473$ K. The inset shows the saturation value
of $\Delta{\cal M}$  as a function of annealing  time (the dashed line
is a guide to the eye).} \label{FIG5}
\end{figure}

As  illustrated in  Fig.  \ref{FIG3}(a),  magnetization curves  show a
significant change  $\Delta{\cal M}$ at  the MT ($\Delta {\cal  M}$ is
estimated   as  the   magnetization  difference   at   the  transition
temperature between  extrapolations of the linear  behaviour of ${\cal
M}$  versus $T$  curves well  above  and well  below the  transition).
Fig.\ref{FIG5}  gives  $\Delta  {\cal  M}$  versus  ${\cal  H}$  after
different annealing times at $T_a$.  In the as-quenched state, $\Delta
{\cal M}$  saturates for a  value of ${\cal  H}$ close to 20  kOe; the
field needed  to reach saturation decreases  with increasing annealing
time. In all  cases, for these values of  the field, the magnetization
itself  is   far  from  having  reached   saturation.   Moreover,  the
saturation  value  of  $\Delta{\cal  M}$ ($\Delta{\cal  M}_{sat}$)  is
strongly dependent on the  annealing time. Actually, it decreases with
$t_a$  as shown  in the  inset of  Fig.  \ref{FIG5},  reaching, within
error, a constant value for $t_a >$ 150 min.

\begin{figure}[t]
\begin{center}
\epsfig{file=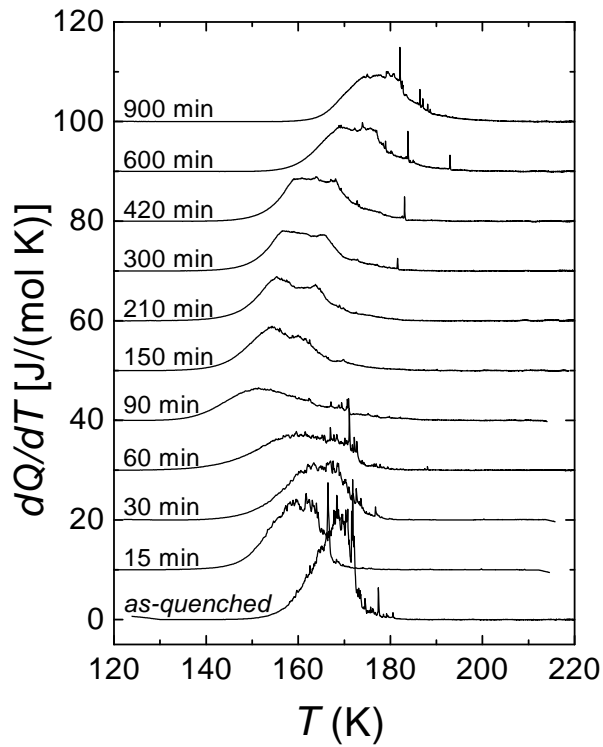, width=8cm}
\end{center}
\caption{Thermal curves corresponding to the reverse (heating)
martensitic transition obtained after different annealing times at
$T_a = 473$ K.} \label{FIG6}
\end{figure}

\begin{figure}[t]
\begin{center}
\epsfig{file=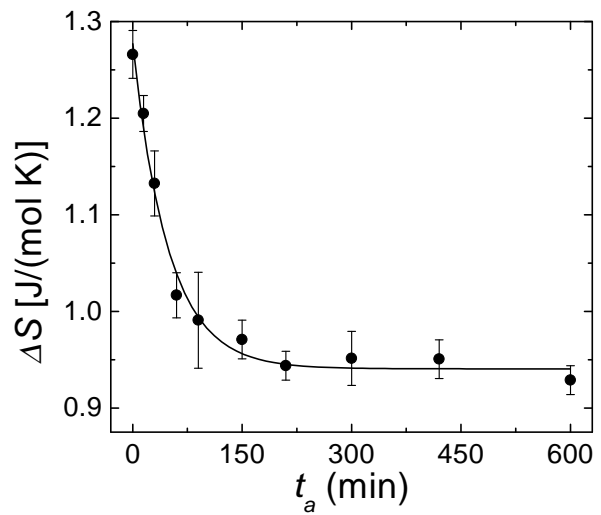, width=8cm}
\end{center}
\caption{Entropy change between $M$-  and $\beta$-phases as a function
of ageing  time at $T_a  = 473$ K.   The solid line corresponds  to an
exponential decay fitted function and serves as a visual guide.}
\label{FIG7}
\end{figure}

As  the $MT$  is  first order,  any  change of  the temperature  $T_M$
induced by the  applied magnetic field ${\cal H}$  should be accounted
for by the Clausius-Clapeyron equation
\begin{equation}
\frac{dT_M}{d {\cal H}}=- \frac{\Delta{\cal M}}{\Delta S},
\end{equation}
where  $\Delta   S$  is  the  entropy  difference   between  $M$-  and
$\beta$-phases.   We  measured  $\Delta  S$  using  calorimetry  after
different annealing  times.  In  Fig.  \ref{FIG6}, we  present thermal
curves  obtained during  heating runs  for different  annealing times.
These curves  show the characteristic noisy structure  which is common
to  many  martensitic transitions  \cite{Clapp,Levin}.   Such a  noisy
structure is enhanced in the as-quenched state.  As the annealing time
increases, the transformation first  becomes smoother and broader, and
shifts towards lower temperatures.   However, for times longer than 90
min thermal curves partially  recover the original noisy character and
the  transformation shifts towards  higher temperatures.   The entropy
change is determined by a numeric integration of the thermograms \cite
{Ortin88}.  The value obtained  for the as-quenched system (1.27 $\pm$
0.02  J/Kmol) is consistent with that reported  in \cite{Prado95}. The
dependence  of $\Delta S$  on  $t_a$  is depicted in Fig.  \ref{FIG7}.
The  figure  reveals  that $\Delta S$  (absolute value) decreases when
the annealing  time  at  $T_a$  increases,  reaching  (as  occurs with
$\Delta {\cal M}_{sat}$)  a  constant  value for $t_a >$ 150 min.  The
relative decrease of $\Delta S$ is about 30$\%$, much smaller than the
relative  decrease  $\Delta {\cal M}_{sat}$, which  is estimated to be
approximately 90$\%$.

It  is  then interesting  to  compare  the  values of  the  derivative
$dT_M/d{\cal  H}$ in  the  as-quenched and  long-term annealed  states
using equation 1. For the as-quenched system, a value of ($8 \pm 0.4$)
mK/kOe  is obtained,  while this  value reduces  to ($1.10  \pm 0.05$)
mK/kOe after 300 min of annealing.  The reduction is mainly related to
the decrease  of $\Delta {\cal  M}_{sat}$ with ageing.   Actually, the
derivative  $dT_M/d{\cal H}$  provides  a good  quantification of  the
dependence  of the  structural  transition temperature  on a  magnetic
field under  the assumption that  $\Delta S$ is independent  of ${\cal
H}$.   Such  an assumption  is  based on  the  fact  that the  entropy
difference between  the open  and close-packed phases  originates from
the  change of the  corresponding vibrational  spectrum, as  occurs in
non-magnetic Cu-based shape-memory alloys \cite{Planes01}.

\begin{figure}[t]
\begin{center}
\epsfig{file=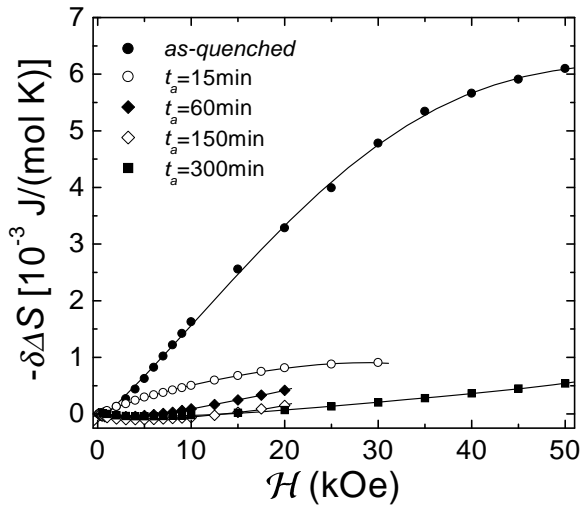, width=8cm}
\end{center}
\caption{Change  in the  entropy  difference between $\beta$ and  $M$
phases $-\delta \Delta S$, as  a function of the magnetic field ${\cal
H}$  after different  annealing  times at  $T_a = 473$  K.}
\label{FIG8}
\end{figure}

A possible magnetic  contribution to $\Delta S$ can  be evaluated from
the  magnetization  vs.  temperature  curves.  Actually,  at  a  given
temperature $T$, the change of entropy of the system with the magnetic
field ${\cal H}$ is given by:
\begin{equation}
\delta S(T, {\cal H}) \equiv  S(T,{\cal H}) -  S(T,{\cal H}=0) =
\int_0^{\cal H}
\left( \frac{\partial {\cal M}}{\partial T}\right)_{\cal H} d{\cal H},
\end{equation}
where the thermodynamic (Maxwell) relation:
\begin{equation}
\left( \frac{\partial S}{\partial {\cal H}}\right)_T =
\left( \frac{\partial {\cal M}}{\partial T}\right)_{\cal H},
\end{equation}
has been  used.  Therefore, application  of a magnetic  field modifies
the entropy change between both  the $M$- and $\beta$-phases according
to:
\begin{equation}
\begin{array}{ccl}
\vspace{0.2cm}
\delta \Delta S & = &  \Delta S ({\cal H}) - \Delta S ({\cal H}=0) \\
 & = &
\displaystyle
\int_0^{\cal H}
\left [ \left( \frac{\partial {\cal M}^M}{\partial T}\right)_{\cal H}
-  \left( \frac{\partial {\cal M}^{\beta}}{\partial T}\right)_{\cal H}
\right ] d{\cal H},
\end{array}
\end{equation}
where $\left( \frac{\partial  {\cal M}^M}{\partial T}\right)_{\cal H}$
and $\left(  \frac{\partial {\cal M}^{\beta}}{\partial T}\right)_{\cal
H}$  are the  derivatives of  the  magnetization with  respect to  the
temperature in the $M$- and $\beta$-phases. The preceding integral was
calculated numerically  from the magnetization  curves above following
the  procedure given  in \cite{Pecharsky99}.   Results are  plotted in
Fig.\ref{FIG8}  as a function  of ${\cal  H}$ for  different annealing
times.  This quantity is always  negative, which means that the effect
of  the  field  is to  increase  the  absolute  value of  $\Delta  S$.
However, the  increase is  very small, and  represents, in  all cases,
less  than 1$\%$ of  the total  entropy change.   It can  therefore be
considered  negligible  for   practical  purposes.   It  is,  however,
interesting to compare this entropy difference with that reported in a
Ni$_{51.5}$Mn$_{22.7}$Ga$_{25.8}$         polycrystalline        alloy
\cite{Hu00}. In that  case a value of $\delta \Delta  S \sim -4 \times
10^{-3}$ J/K  mol is  estimated at  ${\cal H}= 0.9$  T. This  value is
comparable  with the  value  reported here  for  the Cu-Al-Mn  system.
Notice that for the Ni-Mn-Ga system, the change in MT temperature with
magnetic field is also very weak \cite{Planes01}.

\begin{figure}[t]
\begin{center}
\epsfig{file=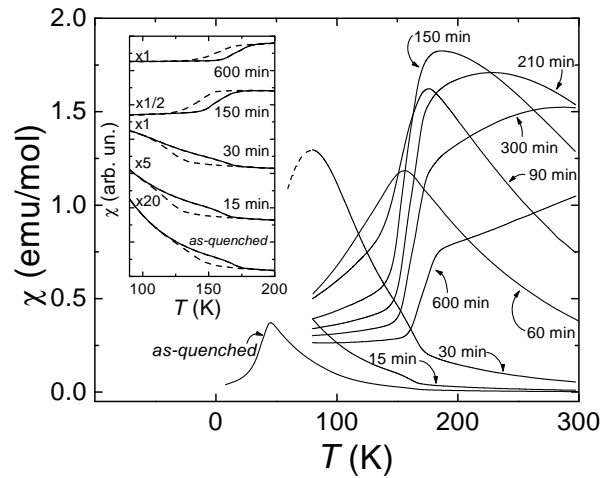, width=8cm}
\end{center}
\caption{Evolution of the real  part of the ac magnetic susceptibility
$\chi$  with annealing  times at  $T_a  = 473$  K. The inset shows  a
detailed  view  of  the   temperature  region  where the  martensitic
transition takes place. Both  heating (solid line) and cooling (dashed
line)  runs are  displayed, revealing  the transformation hysteresis.
For  clarity, the  curves  have been magnified  by a  factor that  is
indicated on each curve.} \label{FIG9}
\end{figure}

An additional, complementary  magnetic characterization of the studied
system  consisted of  measuring the  ac magnetic  susceptibility  as a
function of the  ageing time $t_a$. The evolution of  the real part of
the  ac-susceptibility  is  displayed  in Fig.   \ref{FIG9}.   In  the
as-quenched state,  the magnetic susceptibility  is small and  shows a
low-temperature  peak  associated  with  the appearence  of  a  glassy
magnetic  phase. The martensitic  transition can  be detected  on this
curve as a very  small jump between 160 K and 170  K. The inset of the
figure shows a detailed view of the region where the $MT$ takes place.
In  this case,  heating and  cooling runs  are presented  so  that the
transformation hysteresis is revealed.

As ageing  time  is  increased, the susceptibility peak  moves towards
higher  temperatures  and  becomes  simultaneously higher and broader.
Besides, the evolution  of  the   martensitic  transition  temperature
with  $t_a$  follows  that   of  the  evolution  determined  from  the
calorimetric  measurements   presented  above.  For  annealing   times
$t_a >$ 60 min the  anomalies associated with the freezing temperature
and the martensitic (jump) phases  overlap.  This fact gives rise to a
susceptibility that decreases  abruptly at an intermediate temperature
($\sim$ 150 K) as temperature  is reduced. With further annealing, the
shape of the susceptibility curves remains unchanged, but the value of
$\chi$  decreases  at high  temperatures  indicating  that the  system
becomes more and more ferromagnetic.

\section{DISCUSSION}
\label{discussion}

In  this paper,  we have  reported  results showing  the existence  of
magnetoresistive  effects in  bulk  polycrystalline Cu-Al-Mn  samples.
Magnetoresistive behaviour had  already been observed in $\beta$-phase
melt-spun  ribbons  of similar  composition  \cite{Kainuma98}. In  the
present paper, the investigation has been extended to  the bulk system
and to  the study of  magnetoresistance in the martensitic  phase.  In
both  $\beta$-  and  $M$-phases,  the  observed  magnetoresistance  is
typical   of   heterogeneous  alloys  \cite{Berkowitz92}  where
magnetoresistance  originates from  the  spin-dependent scattering  of
conduction electrons on ferromagnetic  clusters embedded in a metallic
non-magnetic  matrix  \cite{Camley89}.   In  the studied  system,  the
ferromagnetic clusters are due to the tendency of Mn-atoms to increase
their number of third-order neighbours  of the same species within the
basic  $bcc$-lattice.  At  present there is substantial experimental
evidence      of      the      existence     of     these     clusters
\cite{Bouchard75,Kainuma98}.  X-ray  experiments  \cite{Obrado98} have
demonstrated  that  quenched  samples  exhibit an $L2_1$ structure. In
this  structure,  the Mn-atoms are preferentially located on the 4b
sites  (Wyckoff  notation)  of  the  cubic  unit cell of the $Fm3m$
structure.  However,  the distribution of Mn atoms in this sublattice
is  not  completely  random.  Actually, there are indications of the
existence  of  correlations among the positions of the Mn-atoms within
the  sublattice  \cite{Prado98}.  While clusters are not strictly well
defined  in  this case, their influence on transport and magnetic
properties  of the system has to be considered. In particular, they
are  responsible for  the superparamagnetic  behaviour of  the studied
alloy  system, as shown in \cite{Obrado99}. The clustering tendency is
enhanced  by  ageing  within  the miscibility gap. Actually, the
influence of ageing on the magnetic properties of Cu-Al-Mn shows up by
the  remarkable effect that ageing has on the $MR$ curves; a decrease
in  the magnitude of $MR$ at low temperatures and a slight increase at
high  temperatures have been observed (see Fig. \ref{FIG1}). These
modifications are due  to an increase in the  magnetic softening which
takes  place after short annealing times. This is clearly a sign of
the increase of the ferromagnetic character of the system owing to the
growth  of  magnetic  clusters.  Such  an  increase  is confirmed by
magnetization         and         susceptibility         measurements.

The scaling  of $MR$ with  ${\cal M}^2$ was  found to be  different in
both the $\beta$- and $M$-phases.   This effect mainly arises from the
different behaviour of the magnetization in the two phases.  Actually,
the ${\cal M}$ vs.  ${\cal H}/T$  curves also show a tendency to group
into  two  different  families   corresponding  to  the  $\beta$-  and
$M$-phases, respectively.  In the  as-quenched state, the $MR$ shows a
fairly good  linear dependence on  ${\cal M}^2$ for both  the $\beta$-
and  $M$-phases.  This  is the  expected behaviour  in the  absence of
magnetic interaction between magnetic clusters \cite{Wang94}. However,
at  high  fields the curves for the $M$-phase show a clear tendency to
deviate from linear behaviour. A similar deviation has previously been
observed in heterogeneous  alloys and it  has been  attributed to the
progressive  field  alignment  of the disordered spins at the
boundaries  of  the magnetic clusters, which causes a large variation
of the $MR$ compared with  the corresponding  change  in magnetization
\cite{Bellouard94}.  Such  an  increase  in the slope of the $MR$ vs.
${\cal  M}^2$  curves is usually associated with the existence of high
values  of  the  high  field susceptibility, as is the case of the
as-quenched  and  short-term  annealed  samples  [see  magnetization
curves   in   Fig.   \ref{FIG3}(b)].  After  long  time  annealing,
ferromagnetism  is responsible for the loss of linearity in the $MR$
versus  ${\cal  M}^2$  curves.  Actually,  the formation of large
ferromagnetic  clusters   whose  magnetization saturates at relatively
low  fields  (see  Fig. \ref{FIG3}(b)) leads to a large change of the
total  magnetization of the sample and a small variation of $MR$ at
low  fields, since the latter originates from the moment  alignment of
the  small  clusters  and/or the spins at the boundaries of the
ferromagnetic    regions,   which   occurs   at   higher   fields.
Consequently,  $MR$ vs. ${\cal M}^2$ shows a quasi-constant regime
that  corresponds  to  the  range  of  fields  in which the magnetic
saturation  of  large ferromagnetic  clusters  takes  place (see  Fig.
\ref{FIG4})  and  a  subsequent  quasi-linear  behaviour  due  to  the
magnetoresistance  associated  with small clusters and boundary spins.

A  noticeable  effect  of   the  magnetic  field  on  the  martensitic
transition is  the reduction  of the temperature  range $\Delta  T$ in
which the  transition occurs.   A qualitatively comparable  effect has
been   reported  in   the   case  of   the  ferromagnetic   Ni$_2$MnGa
\cite{Ullakko96},  but in this material the reduction is much greater;
for that alloy the transition range reduces from about 10 K without an
applied magnetic  field to less  than 2 K  for an applied field  of 10
kOe. In shape-memory alloys,  the transformation is acknowledged to be
thermoelastic,  which means that  the system  needs to  be continously
cooled  down (or  heated  up)  in order  to  increase the  transformed
fraction of the new phase.   The transition is not completed until the
temperature is lowered (increased) below (above) a certain value.  The
free-energy difference between both  phases provides the driving force
for  the transition.   Such a  driving  force is  proportional to  the
temperature difference between  equilibrium and the actual temperature
to a first approximation \cite{equilibrium}.  The path followed by the
system   is   an   optimal   path   in  which  accommodation  of  the
transformation    elastic    strain   is   almost   accomplished.
Therefore,  $\Delta  T$  represents  a  reasonable  measure of the
stored    elastic    energy   during   the   forward   transformation
\cite{Ortin88}.  From  this  point of view, a decrease in $\Delta T$
reflects a lower value  of this stored elastic energy.  We  argue that
the  physical mechanism behind this  effect is that the magnetic field
around  magnetic clusters breaks the degeneracy of the low-temperature
martensitic phase thus favouring the nucleation of these variants with
the  magnetization  easy-axis  in  the  direction  of  the field. In
Cu-Al-Mn,   approximately orthorhombic  martensitic  domains,  with
the $c$-axis oriented  along  the magnetic field, are expected to have
maximum nucleation   probability.   In  the  Ni-Mn-Ga  alloy  such  an
interpretation  is   corroborated  by  the  fact   that  magnetic  and
mechanical energies  required to  induce a single  variant martensitic
structure    from    a     polyvariant    crystal    are    comparable
\cite{O'Handley98}.

An interesting result concerns the magnetization jump occurring at the
martensitic  transition.   This  jump  is positive  (increase  of  the
magnetization  in the  forward  transition from  the  $\beta$- to  the
$M$-phase) in  the as-quenched state even for  small fields.  However,
for  sufficiently  long  annealing   time,  when  the  sample  behaves
ferromagnetically, this jump is negative for small values of the field
and becomes positive above a certain critical field close to the value
for  which $\Delta {\cal  M}$ saturates.   This is  related to  a high
magnetic  moment  of  the  Mn-clusters  and  to  the  strong  magnetic
anisotropy  of the  martensitic  phase.  For  fields  larger than  the
critical  field,  reorientation of  martensitic  variants enables  the
Zeeman energy to be minimized and the jump again becomes positive.

The   behaviour   of    magnetic   susceptibility   provides   further
understanding  of the  origin of  the interplay  between  magnetic and
structural  degrees  of  freedom.    In  the  as-quenched  state,  the
temperature  dependence  of the  ac-susceptibility  is  that which  is
expected  for  a  system  with  magnetic  clusters.  Below  a freezing
temperature  (maximum  of  the $\chi$  versus  $T$  curve)  the  large
magnetic moments of $L2_1$-clusters are frozen along the magnetic easy
axis. When the ageing time is increased, the peak associated with  the
occurrence  of  the  frozen  magnetic  phase  shifts  towards   higher
temperatures  and  becomes  broader. This  is  a  consequence  of  the
sensitivity of the susceptibility to the  size and  shape distribution
of the magnetic clusters. For annealing times longer than  $\sim$  100
min the behaviour  of  the  susceptibility  is  that  expected  for  a
ferromagnetic   system:  it  is  quite  constant   in  both  $M$-  and
$\beta$-phases. In the temperature range where the $MT$  takes  place,
the increase in the ferromagnetic correlation among clusters  explains
the evolution of the shape of the hysteresis cycle of the $MT$ with an
increase of ageing time. For  short  ageing times, $\chi$ increases at
the $MT$ (forward transition) due to the tendency of the magnetic
moments to freeze. In contrast,  when  the system  becomes
ferromagnetic,  the behaviour of  $\chi$ at  the transition is
dominated by  the magnetic anisotropy which is larger in the $M$- than
in the $\beta$-phase.

From  a  general  viewpoint,  the  results  discussed  above  must  be
considered as evidence of the coupling between magnetic and structural
degrees of  freedom occurring  at a mesoscopic  scale, that is  at the
length scale  of the magnetic/martensitic domains.   The coupling also
exists  at a microscopic  level (spin-phonon  coupling) which  is made
clear by the estimated (through Clausius-Clapeyron equation) change of
$T_M$  with ${\cal  H}$. This  coupling is  however very  weak  and is
related to  the small magnetic contribution to  the entropy difference
between the $\beta$- and $M$-phases.  The obtained value is comparable
to  that estimated  for  ferromagnetic Ni-Mn-Ga,  but is  considerably
smaller  than  the  value  ($\sim  0.1$ K/kOe)  reported  for  ferrous
martensitic alloys \cite{Kakeshita85}. Consistently, it has been shown
that application of a magnetic field does not significantly modify the
phonon branches in Ni-Mn-Ga \cite{Manosa01}.

At  this point,  it is  worth noticing  that the  strong  influence of
ageing on  the entropy difference between $\beta$-  and $M$-phases may
not  be related to  an increase  in the  magnetic contribution  to the
entropy. This effect must be  simply ascribed to the fact that Mn-rich
regions  do not  transform martensitically.   The important  change of
$\Delta S$ takes place within  the ageing  time  interval during which
the  system  becomes  ferromagnetic.  Notice that  simultaneously  the
magnetization change  at the transition reaches  its saturation value.
Therefore, any  further evolution of the magnetic  properties (as, for
instance,  revealed by ac-susceptibility  measurements) of  the system
are originated by short-range reordering processes.

To  conclude,  the  results  presented  prove the  existence  of  both
magnetoresistive and magnetoelastic properties in a non-stoichiometric
Heusler  Cu-Al-Mn  alloy  with  a  Mn  content  of  9  at$\%$ and with
shape-memory  properties.  The occurrence of a martensitic transition
has   enabled  the  study  of  the  effect  of  the  crystallographic
structural   change   on   the   above   mentioned   properties.  The
magnetoresistive behaviour has been  shown to be different for the two
structures.  With  regards  to  magnetoelastic coupling, it has been
found that it mainly occurs  at a mesoscopic level between martensitic
domains               and              magnetic              clusters.

\acknowledgements

This  work has  received  financial support  from  the CICyT  (Spain),
projects  MAT2001-3251,  MAT2000-0858   and  from  CIRIT  (Catalonia),
project  2001SGR00066.  J.M.   is supported  by Direcci\'o  General de
Recerca (Generalitat de Catalunya).



\begin{references}
\bibitem{equilibrium}  The equilibrium temperature  is defined  as the
non-constrained   co-existance  temperature   of   the  $\beta$-   and
$M$-phases.  Since hysteresis is small  in such a class of transition,
this  temperature is  close to  the temperature  of the  onset  of the
forward transition and end of the reverse transition.
\end{references}
\end{document}